\documentclass[twocolumn,prb,english]{revtex4-2}
\usepackage[T1]{fontenc}
\usepackage[latin9]{inputenc}
\usepackage{array}
\usepackage{verbatim}
\usepackage{amsmath}
\usepackage{amssymb}
\usepackage{graphicx}
\usepackage{hyperref}
\usepackage{xcolor}

\makeatletter

\def\phimat{\boldsymbol{\Phi}}
\def\epsmat{\boldsymbol{\epsilon}}
\def\kvec{\boldsymbol{k}}

\let\oldAA\AA
\renewcommand{\AA}{\text{\normalfont\oldAA}}

\global\long\def\ket#1{\left|#1\right\rangle }%

\global\long\def\bra#1{\left\langle #1\right|}%

\makeatother
\begin{document}

\title{Two-Dimensional Spectroscopy of Two-Dimensional Materials}
\begin{abstract}
In this work we provide an exact and efficient numerical approach to simulate multi-time correlation functions in the Mahan-Nozi\`{e}res-De Dominicis model, which crudely mimics the spectral properties of doped two-dimensional semiconductors such as monolayer transition metal dichalcogenides.  We apply this approach to study the coherent two-dimensional electronic spectra of the model. We show that several experimentally observed phenomena, such as peak asymmetry and coherent oscillations in the waiting-time dependence of the trion-exciton cross peaks of the two-dimensional rephasing spectrum, emerge naturally in our approach.  Additional features are also present which find no correspondence with experimentally expected behavior. We trace these features to the infinite hole mass property of the model.  We use this understanding to construct an efficient approach which filters out configurations associated with the lack of exciton recoil, enabling the connection to previous work and providing a route to the construction of realistic two-dimensional spectra over a broad doping range in two-dimensional semiconductors.
\end{abstract}
\author{Lachlan P Lindoy}
\affiliation{Department of Chemistry, Columbia University, 3000 Broadway, New York, New York 10027, USA}
\author{Yao-Wen Chang}
\affiliation{Physics Division, National Center for Theoretical Sciences, Taipei 10617, Taiwan}
\author{David R Reichman}
\email{drr2103@columbia.edu}
\affiliation{Department of Chemistry, Columbia University, 3000 Broadway, New York, New York 10027, USA}

\maketitle

\section{Introduction}
Over the last decade the study of two-dimensional semiconductors and heterostructures constructed from them has greatly enhanced our understanding of fundamental quasiparticle excitations such as excitons, trions and biexcitons, as well as opened a path towards the construction of novel optoelectronic devices~\cite{Akinwande2014,Jariwala2014,Liu2016}.
Among this class of materials, transition metal dichalcogenides (TMDCs) stand out due to the ability to reliably create and manipulate them, as well as the fact that their unique physical properties, such as large spin-orbit coupling and distinguished spin and valley degrees of freedom, create an ideal playground for the observation of novel physical phenomena\cite{doi:10.1021/nl903868w,PhysRevLett.105.136805,Mak2012,Mak2013,You2015,RevModPhys.90.021001,doi:10.1146/annurev-conmatphys-033117-054009}. 
Indeed, a variety of experimental approaches have been employed to reveal behavior ranging from the individual properties of dark excitons to collective behavior such as Wigner crystallization and exciton condensation~\cite{Zhou2017, Fogler2014,Wang2019,Li2021,Zhou2021,Smolenski2021}.

One such experimental technique that has been applied to the study of TMDCs is two-dimensional (2D) electronic spectroscopy~\cite{doi:10.1021/acs.nanolett.6b02041, doi:10.1021/acs.nanolett.1c01098,PhysRevB.104.L241302}. In principle, 2D spectroscopy has the potential to uncover properties such as coherent coupling between excitations that cannot be revealed via linear optical techniques such as absorption or photoluminescence. On the other hand, 2D spectroscopy can be difficult to perform and interpret, and often requires complimentary theory to unlock its full potential.  In the latter regard stands the important work of Tempelaar and Berkelbach, who presented a detailed, microscopic theory of 2D spectra in TMDCs~\cite{Tempelaar2019}.  One possible limitation of their approach, however, is the fact that it is confined to a regime of relatively low electron doping.  Recent work has emphasized the important role played by collective dressing of excitons by excess electrons in the higher doping regime, where the trion peak may be interpreted as emerging from a lower energy branch of a collective exciton-polaron spectrum~\cite{PhysRevB.95.035417,Sidler2017}.

The work presented here aims to enhance our understanding of the emergent features in the 2D spectra of 2D semiconductors by considering a simple model for which exact simulation of the various 2D signals is possible~\cite{PhysRevB.89.245301,PhysRevB.91.115313}.  The downside of such an approach is that the model is less realistic than that studied in Ref. \onlinecite{Tempelaar2019}, but the advantage is that the calculations are non-perturbative and thus can describe the evolution over the full doping range.  Interestingly, we find that several features that arise from our calculations agree with expectations from experiments and past work, while some do not.  A closer examination of the spectral features which appear unique to our approach enable the facile elimination of these features and afford a direct link to the work of Ref. \onlinecite{Tempelaar2019}. This bridge between our phenomenological approach and past microscopic ones should enable the future construction of theoretical tools for the description of non-linear 2D spectra that contain microscopic information but are still capable of capturing non-perturbative effects at high doping~\cite{PhysRevB.99.125421}.

Our paper is organized as follows.  In Sec. \ref{Sec:sec2} the theory of 2D coherent electron spectroscopy for a simple model of 2D semiconductors, a Mahan-Nozi\`{e}res-De Dominicis Hamiltonian based model of electron-exciton
scattering, is presented.  In Sec. \ref{Sec:sec3} the 2D rephasing spectra obtained from this theory are presented, the various features observed are compared against previous theoretical and experimental spectra, and the origin of the features present in the spectra are discussed.  Sec \ref{Sec:sec4} provides a summary and outlook.

\section{2D Spectroscopy for 2D Semiconductors\label{Sec:sec2}}
\subsection{The Mahan-Nozi\`{e}res-De Dominicis Hamiltonian for Electron-Exciton Scattering}

We will consider an idealized model for describing electron-exciton scattering in electron-doped semiconductors.  We take an electron-exciton scattering Hamiltonian of the form of the Mahan-Nozi\`{e}res-De Dominicis (MND) Hamiltonian~\cite{PhysRev.178.1097,JPhysFracne.10.1051,PhysRevB.99.125421}
\begin{equation}
\hat{H} = \mathcal{E} \hat{X}^\dagger\hat{X} + \displaystyle\sum_{\kvec} \epsilon_{\kvec} \hat{c}_{\kvec}^\dagger \hat{c}_{\kvec} + \displaystyle\sum_{\kvec\kvec'} V_{\kvec\kvec'}\hat{c}_{\kvec}^\dagger \hat{c}_{\kvec'}\hat{X}^\dagger\hat{X}. \label{eq:MND_model}
\end{equation}
Here $\hat{X}^\dagger$ and $\hat{X}$ are creation and annihilation operators for an immobile exciton, $\hat{c}_{\kvec}^\dagger$ and $\hat{c}_{\kvec}$ are creation and annihilation operators for an electron in the conduction band with quasi-momentum $\kvec$, $\mathcal{E}$ is the exciton transition energy, $\epsilon_{\kvec} = |\kvec|^2/2m_e$ is the electron kinetic energy with $m_e$ the electron mass, and $V_{\kvec\kvec'}$ the scattering potential.  

Assuming a 1S exciton state, we take the scattering potential to be~\cite{PhysRevB.99.125421}
\begin{equation}
V_{\kvec\kvec'} = \frac{-\nu_{\kvec-\kvec'}}{A} \left(1 - \exp\left[-\frac{1}{2}|\kvec-\kvec'|^2 \zeta^2 \right]\right),
\end{equation}
where $\zeta$ is the exciton radius, $A$ is the area, and $\nu_{\kvec-\kvec'}$ is the screened Coulomb potential and here is taken to be the Rytova-Keldysh potential\cite{ rytova1967, keldysh_pot, PhysRevB.88.045318}
\begin{equation}
\nu_{\kvec-\kvec'} = \frac{2\pi e^2}{|\kvec-\kvec'|(1+r_0|\kvec-\kvec'|)},
\end{equation}
where $r_0$ is the screening length.  We note that this form of the potential is microscopically derived in the limit of infinite exciton effective mass, which is the correct limit for the MND model~\cite{doi:10.1063/5.0008730}.  If, however, we endeavor to more realistically treat the nearly equal electron and hole masses as found in TMDCs, a different effective potential should be employed~\cite{PhysRevB.103.075417}.

Within this model, electron doping may be included by considering an initial electron distribution ($n_{\kvec}$) given by a 2D non-interacting electron gas with
\begin{equation}
n_{\kvec} = \frac{1}{\exp \left[ \beta \left( \epsilon_{\kvec} - \epsilon_f\right)\right] + 1},
\end{equation}
where $\epsilon_f$ is the Fermi energy, and $\beta$ the inverse temperature. In 
the zero temperature limit, this reduces to 
\begin{equation}
n_{\kvec} = 1-\Theta(\epsilon_{\kvec}-\epsilon_f).
\end{equation}

This model ignores spin degrees of freedom, and exchange interactions, and as mentioned above, assumes an immobile (infinite mass) exciton with a purely 1s character that couples to a non-interacting Fermi sea.  As such, it is not expected to quantitatively describe experimental results. However, this model is able to qualitatively capture the emergence of a trion peak, oscillator strength transfer, and doping-dependent line shapes in the linear absorption spectrum of monolayer TMDCs even at relatively high electron doping~\cite{PhysRevB.99.125421}.  Here we employ this simplified model to gain insight into the non-linear 2D spectrum of 2D materials such as TMDCs, including its doping dependence.  While limited, the depth with which we are able to investigate physical features enables an understanding of the more realistic microscopic situation probed in real experiments.

\subsection{Multi-time Correlation Functions}
2D coherent spectroscopy provides a means of monitoring coherent and incoherent dynamics of a system.  This technique employs a sequence of three ultrafast pulses in order to produce a nonlinear optical response in the material~\cite{mukamel, doi:10.1021/acs.nanolett.6b02041}. By measuring this response and its dependence on the timing between pulses, it is possible to resolve couplings between distinct states in the system, and to differentiate between coherent and incoherent energy transfer processes~\cite{doi:10.1021/acs.nanolett.6b02041,https://doi.org/10.1002/lpor.201800171}.

\begin{figure}[h!]
\begin{centering}
\includegraphics[width=\columnwidth]{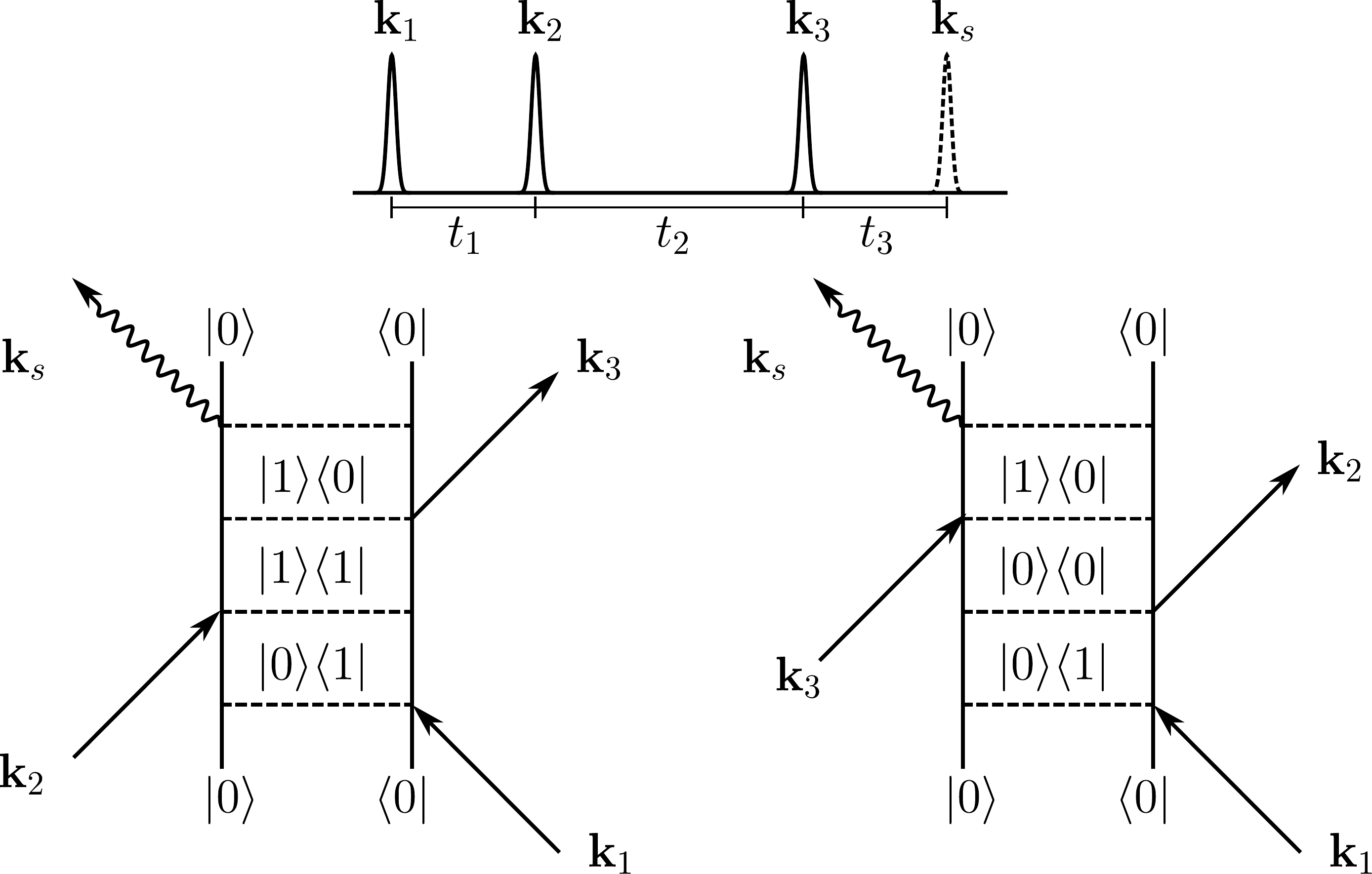}
\par\end{centering}
\caption{\label{fig:fig1} Illustration of the pulse time ordering used for obtaining 2D rephasing spectra (top).  The double-sided Feynman diagrams representing the Liouville space pathways that contribute to the 2D rephasing spectrum (bottom).   These contributions are accounted for by the correlation functions $R_2(t_1, t_2, t_3)$ (left) and $R_3(t_1, t_2, t_3)$ (right), and correspond to stimulated emission and ground state bleach spectra, respectively. Here we have used $\ket{0}$ and $\ket{1}$ to represent zero and one exciton states, respectively, and at each stage arbitrary configurations of the conduction band electrons are allowed.}
\end{figure}

For the electron-immobile exciton scattering model we consider here, the light matter interaction operator takes the form~\cite{PhysRevB.99.125421}
\begin{equation}
\hat{V} = c^*\hat{X}^\dagger + c\hat{X},
\end{equation}
where $c$ is a scalar proportional to the transition momentum matrix element.  
With this definition, and within the rotating wave approximation, a total of four distinct multi-time correlation functions (and their complex conjugates) can provide significant contributions to the nonlinear optical response~\cite{mukamel}.  These are the non-rephasing stimulated emission, $R_1$, the rephasing stimulated emission, $R_2$, the rephasing ground state bleach, $R_3$, and the non-rephasing ground state bleach, $R_4$, contributions. Here, for simplicity and in order to allow for qualitative comparisons with previously obtained experiments~\cite{doi:10.1021/acs.nanolett.6b02041}, we only consider the two rephasing contributions which are given by
\begin{align}
R_2(t_1,\! t_2,\! t_3) &= \mathrm{Tr}\!\left[ \hat{V}(t_1\!+\!t_2) \hat{V}(t_1\!+\!t_2\!+\!t_3)\hat{V}(t_1) \hat{\rho} \hat{V}\right], \label{eq:R2} \\
R_3(t_1,\! t_2,\! t_3) &= \mathrm{Tr}\!\left[ \hat{V}(t_1) \hat{V}(t_1\!+\!t_2\!+\!t_3)\hat{V}(t_1\!+\!t_2) \hat{\rho} \hat{V}\right] \label{eq:R3}. 
\end{align}
The corresponding 2D rephasing spectrum can be obtained from this correlation functions as
\begin{equation}
    S_{RP}(\omega_1, t_2, \omega_3) = S_2(\omega_1, t_2, \omega_3) + S_3(\omega_1, t_2, \omega_3), \label{eq:rephasing_spectrum}
\end{equation}
where 
\begin{equation}
    S_{i}(\omega_1, t_2, \omega_3)= \int_0^\infty\!\mathrm{d}t_1e^{-i\omega_1 t_1}\!\!\!\int_0^\infty\!\mathrm{d} t_3e^{i\omega_3 t_3}R_i(t_1,\! t_2,\! t_3),
\end{equation}
for $i=2,3$.  Here $\omega_1$ and $\omega_3$ are the excitation and emission energies, respectively, and $t_2$ is the waiting-time. In Fig. \ref{fig:fig1}, we illustrate the quantum mechanical pathways that contribute to this spectrum.

\subsection{2D Spectroscopy of the MND model}

If we assume that the system is initially in a state with no exciton and with an initial thermal (or ground state if at zero temperature) configuration of the conduction electrons, it becomes possible to evaluate the multi-time correlation functions (given in Eqs. \ref{eq:R2} and \ref{eq:R3}) exactly for the electron-exciton scattering form of the MND model. A given multi-time correlation function can be expressed as
\begin{align}
R_i(t_1, t_2, t_3) = \mathrm{det}\left[(\boldsymbol{1} - \boldsymbol{n}) + \boldsymbol{n} \boldsymbol{R}_i(t_1, t_2, t_3) \right], \label{eq:Ri_det_T}
\end{align}
where $\boldsymbol{1}$ is the identity matrix, and $\boldsymbol{n}$ is the matrix with elements $[\boldsymbol{n}]_{\kvec\kvec'} = \delta_{\kvec\kvec'} n_{\kvec}$.  
Restricting ourselves to the rephasing contributions, we have
\begin{align}
\boldsymbol{R}_2(t_1,\!t_2,\!t_3) &=\phimat e^{i \tilde{\epsmat} (t_1+t_2)} \phimat^\dagger e^{i\epsmat t_3}\phimat e^{-i\tilde{\epsmat}(t_2+t_3)}\phimat^\dagger e^{-i\epsmat t_1},\label{eq:r2_matrix} \\
\boldsymbol{R}_3(t_1,\!t_2,\!t_3) &= \phimat e^{i \tilde{\epsmat} t_1} \phimat^\dagger e^{i\epsmat (t_2+t_3)}\phimat e^{-i\tilde{\epsmat}t_3}\phimat^\dagger e^{-i\epsmat (t_1+t_2)} 
\label{eq:r3_matrix}, 
\end{align}
where $\epsmat$ is a matrix with elements $[\epsmat]_{\kvec\kvec'} = \delta_{\kvec\kvec'} \epsilon_{\kvec}$, and $\phimat$ and $\tilde{\epsmat}$ are obtained by solving the eigenvalue problem
\begin{equation}
\displaystyle\sum_{\kvec\kvec'} \left((\epsilon_{\kvec} + \mathcal{E})\delta_{\kvec\kvec'} + V_{\kvec\kvec'}\right) \Phi_{\boldsymbol{k}'n} = \Phi_{\boldsymbol{k}n}\tilde{\epsilon}_n.
\end{equation}
In the zero temperature limit, Eq. \ref{eq:Ri_det_T} reduces to
\begin{equation}
R_i(t_1, t_2, t_3) = \mathrm{det}\left[\boldsymbol{R}_i(t_1, t_2,t_3)\right]_{\epsilon_{\kvec} \leq \epsilon_f}, \label{eq:rdet}
\end{equation}
where the determinant is only evaluated over the rows and columns of the matrix $\boldsymbol{R}_i$ with $\epsilon_{\kvec}$ less than or equal to the Fermi energy. 
Similar expressions can be obtained for the non-rephasing contributions.

\subsection{Computational Details}
Before we can numerical evaluate  Eqs. \ref{eq:r2_matrix}, \ref{eq:r3_matrix}, and \ref{eq:rdet} it is necessary to specify a discretization of the model.  We use a finite-size square box with square lattice points given by
\begin{equation}
    \kvec = (k_x, k_y) = \left(\frac{\pi(2\kappa_x-(N-1))}{L} , \frac{\pi(2\kappa_y-(N-1))}{L} \right),
\end{equation}
where $\kappa_x, \kappa_y = 0, 1, \dots, N-1$.  
Here $N$ is the number of grid points in one direction and $L=Na_\Delta$ is the box length, where $a_\Delta$ is a cutoff length that is set by the choice of a cutoff energy $E_\Delta = \mathrm{max}(k_x^2/2 m_e)$ (here taken to be $E_\Delta = 1$ eV).  We have used $N=140$ grid points per dimension for all calculations presented here.  
\begin{figure*}[t!]
\begin{centering}
\includegraphics[width=\textwidth]{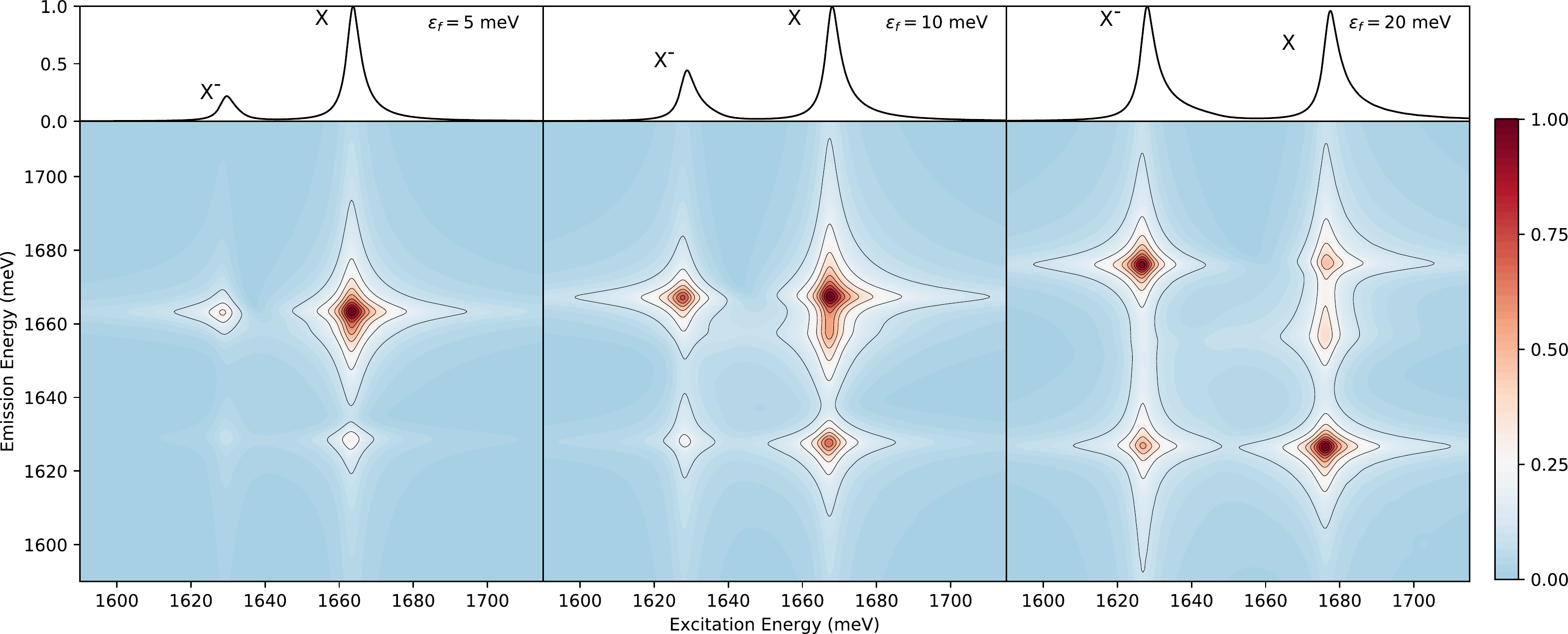}
\par\end{centering}
\caption{\label{fig:fig2} Doping dependence of the linear absorption spectrum, and the amplitude 2D rephasing spectrum of the MND model calculated with a waiting-time of $t_2=0$.  Here we have normalized each of 2D amplitude spectra so that the maximum amplitude is 1.}
\end{figure*}

In our calculations we have chosen the screening length, electron mass, and exciton transition energy consistent with those of MoSe$_2$.  We have taken a screening length of $r_0 = 51.7 \mathrm{\AA}$ (obtained from the 2D polarizability used in Ref. \onlinecite{Tempelaar2019} obtained via the approximate relation $r_0 = 2\pi\chi_{2d}$)~\cite{doi:10.1146/annurev-conmatphys-033117-054009}, electron mass of $m_e = 0.52 m_0$~\cite{Durnev_2018}, and exciton transition energy $\mathcal{E}=1.66$ eV~\cite{Tempelaar2019}   .
As was the case in Ref. \onlinecite{PhysRevB.99.125421}, the exciton radius, $\zeta = 8 \AA$, was taken as an adjustable parameter that was used to control the trion-binding energy.  
We again emphasize, however, that the form of our exciton-electron scattering potential is only consistent in the limit $m_{ex}\rightarrow\infty$, a situation distinct from that of typical TMDCs~\cite{PhysRevB.103.075417}.

In order to compute the 2D rephasing spectrum (Eq. \ref{eq:rephasing_spectrum}) for a given waiting-time it is necessary to evaluate the two rephasing multi-time correlations function over a dense set of time points ($t_1$, $t_3$).  We have used a $2000\times 2000$ grid of points ($t_1, t_3$) in order to obtain the spectra presented here. For each time-point we need to construct the two ($N^2\times N^2$) dense matrices $\boldsymbol{R}_2$ and $\boldsymbol{R}_3$ for which we need to evaluate the determinant.  For the large matrices and grid sizes considered here this process can be rather costly, however, it can be rendered significantly more efficient via GPU acceleration. All results presented here were obtained using a GPU implementation of Eqs. \ref{eq:r2_matrix}, \ref{eq:r3_matrix}, and \ref{eq:rdet}.

\section{2D Rephasing Spectra \label{Sec:sec3}}
\subsection{Doping-Dependent 2D Rephasing Spectra}

In Fig. \ref{fig:fig2}, we present the linear absorption spectra and absolute value of the 2D rephasing spectra obtained with a waiting-time $t_2=0$, at zero temperature, and for varying values of the Fermi energy, $\epsilon_f$.  Consistent with previous experimental and theoretical studies~\cite{doi:10.1021/acs.nanolett.6b02041,Tempelaar2019}, each spectrum shows four peaks arranged in a square pattern, readily attributed to the bound trion and exciton states.  The significant asymmetry that is observed in the line shapes of the linear absorption spectra~\cite{PhysRevB.99.125421} is evident in the 2D spectra as the long tails of the peaks (most evident in the exciton-exciton peak).

As the Fermi energy increases there is considerable oscillator strength transfer from the exciton-exciton peak to the trion-trion peak, consistent with the linear spectra~\cite{PhysRevB.99.125421}.  The two cross-peaks, corresponding to excitation at the exciton(trion) and emission at the trion(exciton) energies which we will refer to as the X-X$^-$(X$^-$-X), are more significant as the Fermi energy increases, becoming the dominant feature at $\epsilon_f = 10$ meV.  Additionally the Fermi energy increases, it becomes possible to resolve an additional peak (that for smaller Fermi energies appears as a shoulder in the exciton-exciton peak).  We discuss the origin of this peak in section \ref{sec:additional_peak}.

\begin{figure*}[t]
\begin{centering}
\includegraphics[width=\textwidth]{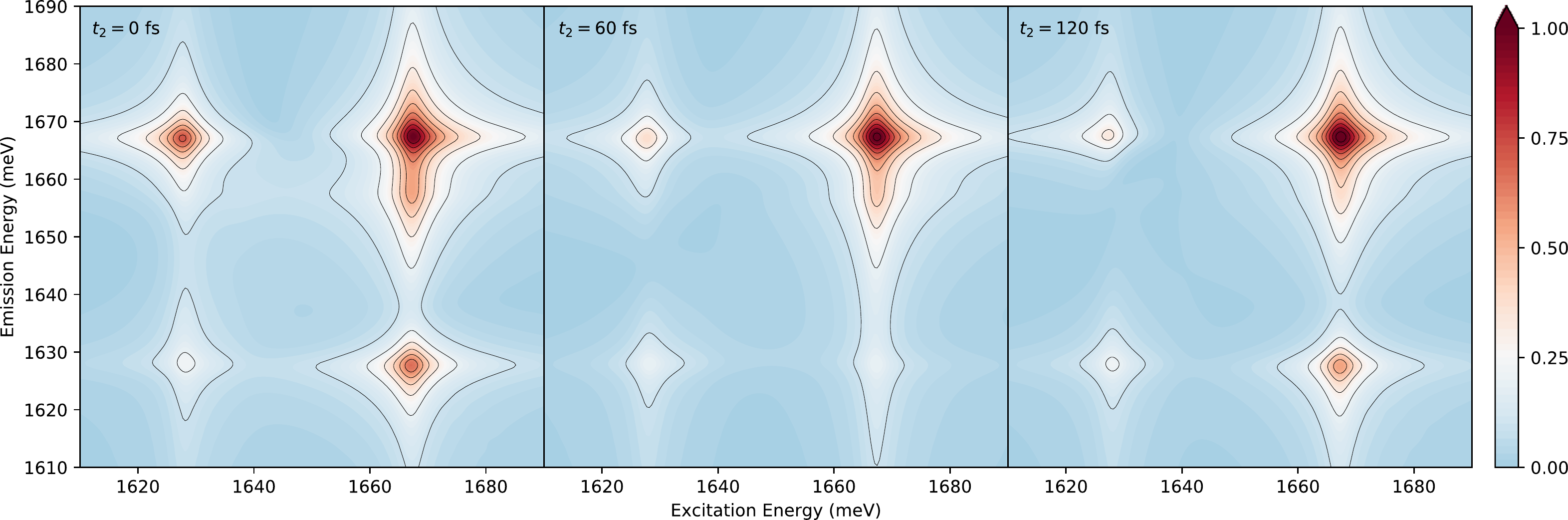}
\par\end{centering}
\caption{\label{fig:fig3} Amplitude 2D rephasing spectrum of the MND model with $\epsilon_f = 10$ meV calculated for a range of waiting-times $t_2$ (shown on each panel).  Here, each of the 2D amplitude spectra have been normalized by the amplitude of the largest peak of the $t_2=0$ spectrum.}
\end{figure*}

\begin{figure}[h]
\begin{centering}
\includegraphics[width=\columnwidth]{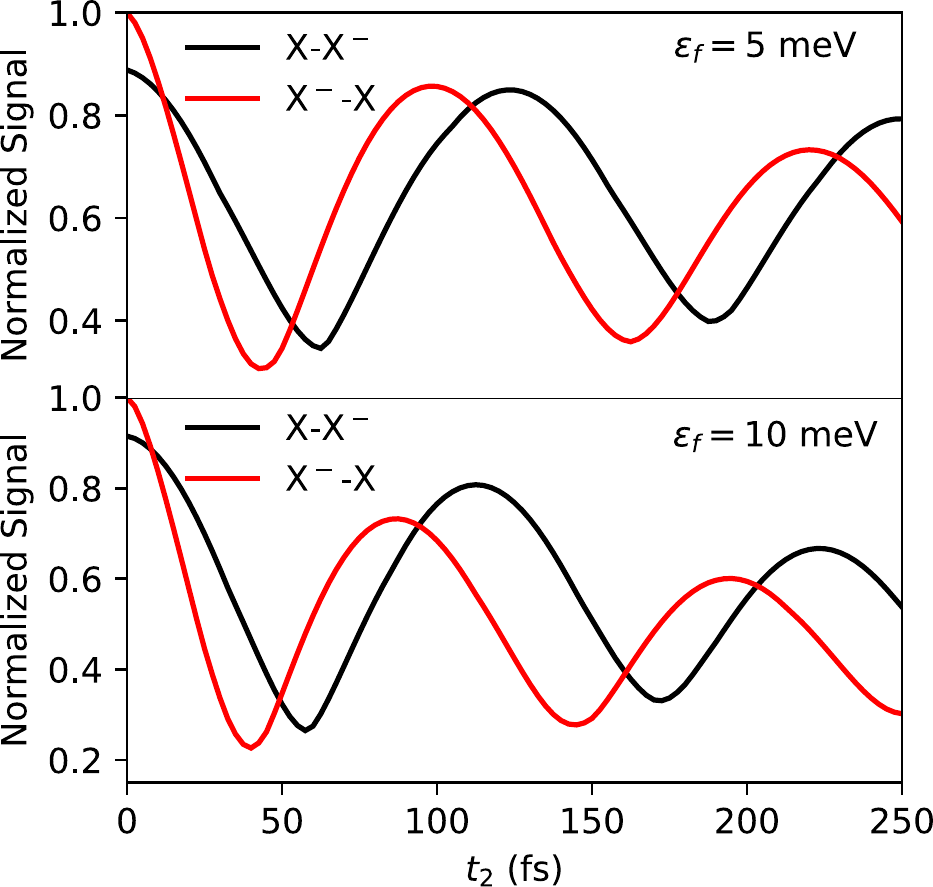}
\par\end{centering}
\caption{\label{fig:fig3b} The waiting-time, $t_2$, dependent normlized amplitude of the 2D rephasing spectrum at the lower (X-X$^-$) and upper (X$^-$-X) cross peaks obtained for the MND model with $\epsilon_f = 5$ (top) and $10$ (bottom) meV.  The two curves in each panel have been normalized by the same constant so that the maximum value obtained by either of the curves is 1.}
\end{figure}

The presence of the cross peaks in the spectra is indicative of exciton-trion coupling. Such peaks can arise from both coherent and incoherent processes.  In order to distinguish between these distinct mechanisms it is necessary to consider 2D spectra evaluated for different values of the waiting-time, $t_2$.  In Fig. \ref{fig:fig3}, we present the absolute value of the 2D rephasing spectrum at varying values $t_2$ for a Fermi energy of $\epsilon_f = 10$ meV. The diagonal peaks remain relatively unchanged.  In contrast, oscillations are observed in the amplitudes of the cross-peaks, with the amplitudes significantly diminished at $t_2=60$ fs compared with either $t_2=0$ or $t_2=120$ fs.  This behavior is qualitatively consistent with the experimental 2D rephasing spectrum obtained for MoSe$_2$~\cite{doi:10.1021/acs.nanolett.6b02041}.
In addition to this, we note that the intensity of the shoulder peak decreases with increasing waiting-time.

\begin{figure*}[t]
\begin{centering}
\includegraphics[width=\textwidth]{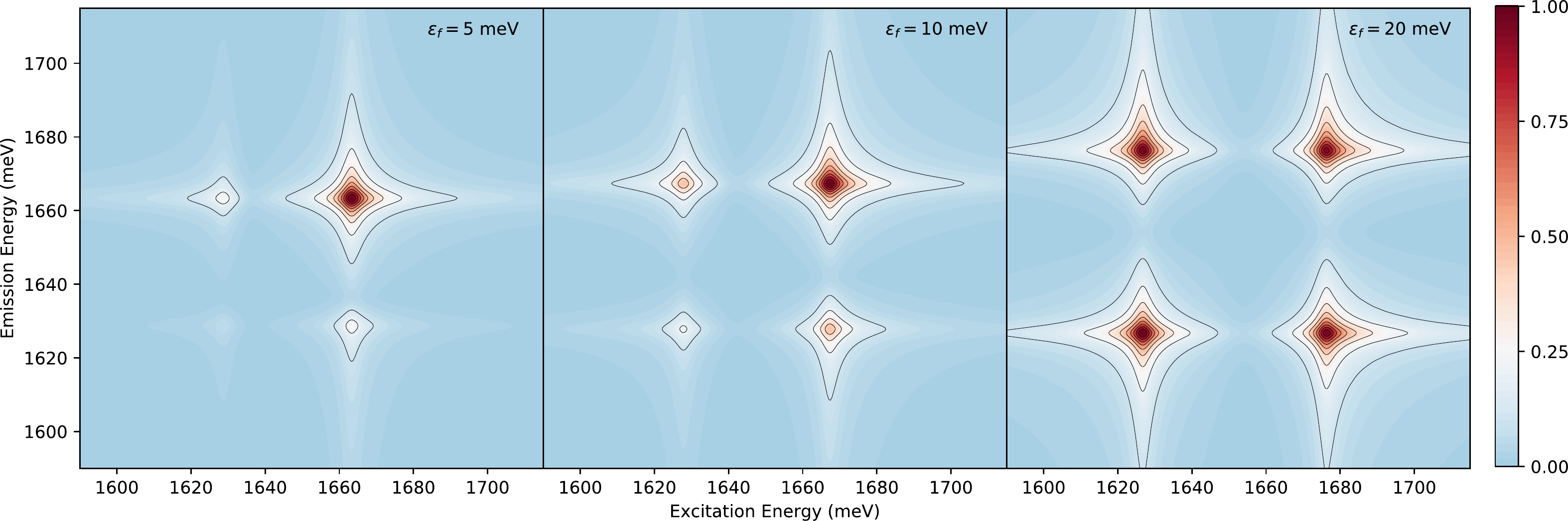}
\par\end{centering}
\caption{\label{fig:fig4} Amplitude 2D rephasing spectrum of the MND model with the final state of the conduction electrons restricted to the ground state configuration at a waiting-time $t_2=0$ and with varying Fermi energies.}
\end{figure*}

The oscillation in the amplitude of the cross peaks is made more explicit in Fig. \ref{fig:fig3b}, where we present the waiting-time dependent amplitude of the X-X$^-$ and X$^-$-X cross peaks for the MND model with $\epsilon_f = 5$ and $10$ meV.  Here, we observe well defined oscillations in the peak amplitudes supporting the fact that the system undergoes coherent interconversion between exciton and trion states, as well as a decay with time, consistent with experimental spectra~\cite{doi:10.1021/acs.nanolett.6b02041}.  The frequency of the oscillation in the cross peaks increases with increasing Fermi energy, $\epsilon_f$, due to the increase in the energy splitting between the exciton peak and the trion peak with increasing doping. Additionally significant asymmetry in the  X-X$^-$ and X$^-$-X peak amplitudes is observed, in contrast to the spectra of Ref. \onlinecite{Tempelaar2019}.  In the next section we will explore the origin of this asymmetry and of the additional peak observed in the spectra.

\subsection{Restricted Final State Spectra \label{sec:additional_peak}}
In addition to the expected four peaks that can be attributed to the bound trion and exciton states, and coherences between them, the 2D rephasing spectra shown in Fig. \ref{fig:fig2} exhibit an additional peak at large doping corresponding to excitation at the exciton energy followed by emission at an energy between the exciton and trion energies.  
The identification of the origin of this peak in the 2D spectra is complicated 
by the large number of distinct Liouville space pathways that contribute to the 2D rephasing spectra.  While the state of the exciton following the application of a pulse is restricted to those shown in Fig. \ref{fig:fig1}, the conduction electrons are free to take any dynamically allowed configuration.

In an effort to obtain a better understanding of the source of the additional peak observed in the 2D rephasing spectra at high Fermi energies, restrict the possible Liouville space pathways by restricting the final configuration of the conduction electrons.
This can be done by replacing the light-matter interaction term at time $t_1+t_2+t_3$ in Eqs. \ref{eq:R2} and \ref{eq:R3} with a new term that involves a projection onto the desired final conduction electron state, $\ket{\Psi_f}$, that is
\begin{equation}
    \hat{V}(t_1\!+\!t_2\!+\!t_3) \rightarrow \hat{V}_f(t_1\!+\!t_2\!+\!t_3)
\end{equation}
with 
\begin{equation}
    \hat{V}_f = \hat{V} \otimes \ket{\Psi_f}\bra{\Psi_f}.
\end{equation}
Here we will note that at $t_2=0$, that the two contributions to the rephasing spectra are equivalent and we thus only provide explicit results for rephasing stimulated emission correlation function, $R_2$:   
\begin{align}
R^{f}_2(t_1,\!t_2,\!t_3)= \mathrm{det}\left[\boldsymbol{M}^{(1)}_{[\boldsymbol{k}_0][\boldsymbol{k}_f]}(t_1,\! t_2,\! t_3) \boldsymbol{M}^{(2)}_{[\boldsymbol{k}_f][\boldsymbol{k}_0]}(t_1,\! t_2,\! t_3)\right], \label{eq:r2bexp}   
\end{align}
where
\begin{align}
\boldsymbol{M}^{(1)}(t_1, t_2, t_3) &= \phimat e^{i \tilde{\epsmat} (t_1+t_2)} \phimat^\dagger e^{i\epsmat t_3}, \\
\boldsymbol{M}^{(2)}(t_1, t_2, t_3) &=  \phimat e^{-i\tilde{\epsmat}(t_2+t_3)}\phimat^\dagger e^{-i\epsmat t_1},
\end{align}
and we have used the notation $\boldsymbol{A}_{[\boldsymbol{a}][\boldsymbol{b}]}$ to denote the matrix formed from the rows $\boldsymbol{a}$ and columns $\boldsymbol{b}$ of the matrix $\boldsymbol{A}$.  Related expressions hold for the other three multi-time correlation functions.

Constraining the final state of the conduction electrons to be the ground state, we can obtain the 2D-rephasing spectrum using $R_2^0$.  At $t_2=0$, this spectrum is identical to the non-rephasing stimulated emission spectrum with the final state constrained to be in the ground state, which is the spectrum that was considered in Ref. \onlinecite{Tempelaar2019}.
In Fig. \ref{fig:fig4}, we present this constrained rephasing spectrum for a waiting-time of $t_2=0$, and for a range of Fermi energies.  As with the full spectrum present in Fig. \ref{fig:fig2}, the asymmetric lineshapes of the one-dimensional spectra are strongly reflected by the shapes of the peaks in the 2D spectra, and these spectra capture the majority of the transfer of oscillator strength from the exciton to trion peak with increasing Fermi energy.  
Additionally, these ground state restricted spectra are symmetric around the line $\omega_1 = \omega_3$, consistent with the results in Ref. \onlinecite{Tempelaar2019}.  This is a consequence of restricting the final state of the bath, it follows immediately from Eq.
\ref{eq:r2bexp}.  As a result the asymmetry present in the full spectra is not observed when the final states of the conduction electrons is constrained to the ground state.  The additional peak (and asymmetry in the cross peaks) arise from pathways in which, following the application of the three pulses and subsequent emission from the sample, there is at least one conduction electron that has been excited out of the Fermi Sea. The remaining question is: Which states give rise to this peak?

Now, it is clearly impractical to perform an exhaustive search of all final states, as there are too many individual states and it is not immediately clear that a contribution from a given state will necessarily correspond to a feature in the spectrum.  As such, we now turn to looking at the contributions to the spectra from pathways ending in states with a specific number of excitations of the Fermi sea.  

For the set of all states that contain a single excitation of a conduction electron from the Fermi sea, the modified light matter interaction that is to be employed is
\begin{equation}
    \hat{V}_f = \hat{V}\otimes \sum_{i\in\boldsymbol{k}_0} \sum_{j\in\setminus\boldsymbol{k}_0}  \hat{c}^\dagger_j\hat{c}_i\ket{\Psi_0}\bra{\Psi_0}\hat{c}_i^\dagger\hat{c}_j,
\end{equation}
where the sum over $i$ runs over the set of all $K$ initially occupied orbitals $\boldsymbol{k}_0$ and $j$ over the set of all unoccupied orbitals, here denoted by $\setminus \boldsymbol{k}_0$. Introducing the notation $\boldsymbol{k}_i^j$ to denote the set of orbitals occupied by $\hat{c}^\dagger_j\hat{c}_i\ket{\Psi_0}$, the singly-excited state restricted multi-time correlation functions may be expressed as
\begin{equation}
\begin{split}
    R_2^{(1)}(t_1, t_2, t_3) = \sum_{i\in\boldsymbol{k}_0} \sum_{j\not\in\boldsymbol{k}_0} &\mathrm{det}\left[\boldsymbol{M}^{(1)}_{[\boldsymbol{k}_0][\boldsymbol{k}_i^j]}(t_1,\! t_2,\! t_3)\right]\\
    \cdot & \mathrm{det}\left[ \boldsymbol{M}^{(2)}_{[\boldsymbol{k}_i^j][\boldsymbol{k}_0]}(t_1,\! t_2,\! t_3).\right]
\end{split}
\end{equation}

Given the potentially large number terms in each of these sums, evaluation of this expression is infeasible.  
Instead, we note that, with some rearrangement and the use of the Cauchy-Binet formula, the sum over $j$ may be performed analytically, giving
\begin{equation}
    R_2^{(1)}(t_1, t_2, t_3) = \sum_{i\in\boldsymbol{k}_0}\left( \mathrm{det}\left[\boldsymbol{R}^{(i)}_{2}(t_1,\! t_2,\! t_3)\right]\right) - KR_2^0(t_1,\!t_2, \!t_3).
\end{equation}
Here $\boldsymbol{R}^{(i)}_{2}(t_1,\!t_2,\!t_3)$ is a $(2K-1)\times(2K-1)$ matrix of the form
\begin{equation}
  \boldsymbol{R}^{(i)}_{2}(t_1,\!t_2,\!t_3) = (-1)^{K-1} \begin{pmatrix}
\boldsymbol{0}_{(K-1) \times (K-1)} & \boldsymbol{A}(t_1,\!t_2,\!t_3) \\
\boldsymbol{B}(t_1,\!t_2,\!t_3) & \boldsymbol{C}(t_1,\!t_2,\!t_3) \end{pmatrix},
\end{equation}
where $\boldsymbol{0}_{(K-1) \times (K-1)}$ is the $(K-1)\times(K-1)$ matrix of zeros, and the matrices $\boldsymbol{A}$, $\boldsymbol{B}$, $\boldsymbol{C}$ are the $(K-1)\times K$, $K\times (K-1)$, and $K\times K$ matrices,
\begin{align}
    \boldsymbol{A}(t_1,\!t_2,\!t_3) &= \boldsymbol{M}^{(2)}_{[\boldsymbol{k}_0\setminus i][\boldsymbol{k}_0]}(t_1,\!t_2,\!t_3), \\
    \boldsymbol{B}(t_1,\!t_2,\!t_3) &= \boldsymbol{M}^{(1)}_{[\boldsymbol{k}_0][\boldsymbol{k}_0\setminus i]}(t_1,\!t_2,\!t_3), \\
    \boldsymbol{C}(t_1,\!t_2,\!t_3) &= \boldsymbol{M}^{(1)}_{[\boldsymbol{k}_0][\setminus\boldsymbol{k}_0 ]}(t_1,\!t_2,\!t_3)\boldsymbol{M}^{(2)}_{[\setminus\boldsymbol{k}_0 ][\boldsymbol{k}_0]}(t_1,\!t_2,\!t_3),
\end{align}
where $[\boldsymbol{k}_0\setminus i]$ denotes the set of $K-1$ indices obtained after removing $i$ from the set $\boldsymbol{k}_0$.

\begin{figure*}[t]
\begin{centering}
\includegraphics[width=\textwidth]{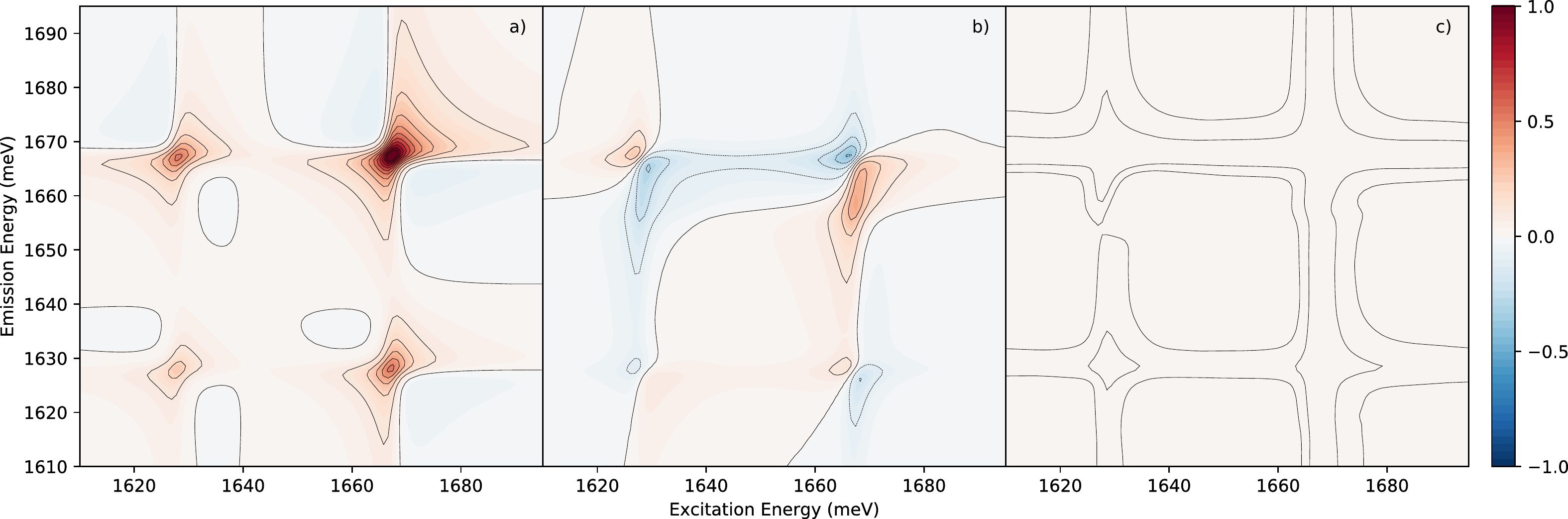}
\par\end{centering}
\caption{\label{fig:fig5} Real part of the 2D rephasing spectrum of the MND model for $\epsilon_f=10$ meV and a waiting-time $t_2=0$ with the final state of the conduction electrons restricted to the ground state configuration a), all singly-excited state configurations b), and all higher excitation configurations c). Here each spectra is normalized by the normalisation constant required to ensure that the peak in the full amplitude spectrum has amplitude 1. Note we have used a finer spacing of contour lines in panel c) in order to better show the structure in the spectrum.}
\end{figure*}
In Fig. \ref{fig:fig5} we compare the real parts of the $t_2=0$ rephasing spectra obtained with the final state constrained to the ground state of the conduction electron bath (a), the set of all singly-excited states (b), and the remaining contributions coming from all final states with more than a single excitation for a system with a Fermi energy of $\epsilon_f = 10$ meV. From panel (a) we observe that the dominant contributions to the peaks arising from the exciton, trion and their coherences can be attributed to pathways in which the conduction electrons return to their final state.  The singly-excited state contributions shown in panel (b) introduce asymmetry to the spectra, providing small contributions to each of the peaks present in the ground state spectra, and additionally introduce the additional peak.  All pathways in which the final state of the conduction electron bath contains more than a single-excitation above the Fermi sea provide very minor changes to the spectra.  

These results demonstrate that the presence of the additional peak and asymmetry in the 2D spectra can entirely be attributed to terms in which, following the three pulse sequence and emission of a photon, the system returns to a state with no exciton, but with a single conduction electron excited above the Fermi sea.  Now such a state will in general have a different linear momentum from the initial configuration of the system, and so for realistic 2D coherent spectroscopy experiments (in which conservation of photon momentum, and therefore the system momentum, is used to resolve different pathways), these states should not contribute to the spectrum.

Within the electron-exciton scattering form of the MND Hamiltonian (Eq. \ref{eq:MND_model}), the presence of such features is not surprising.  In arriving at this model, it is necessary to take the infinite exciton mass limit.  Upon doing so the original scattering model becomes an impurity model that does not preserve linear momentum during interactions between the conduction electrons and exciton.  As such, we can view this feature as an artifact of the immobile exciton model.  It is important to note that, consistent with the above discussion, the physically filtered case (Fig \ref{fig:fig5}a)) looks nearly identical to the spectrum obtained in a different manner in Ref. \onlinecite{Tempelaar2019}, where the mass of the exciton is finite.

\section{Conclusion \label{Sec:sec4}}
In this paper, we have applied a Mahan-Nozi\`{e}res-De Dominicis Hamiltonian based model of electron-exciton scattering to the evaluation of 2D spectroscopy for 2D materials.  This simple model, in which we assume that the exciton is immobile, qualitatively captures many of the features that have been observed in experimental and previous theoretical treatments of the multi-dimensional spectroscopy of monolayer transition metal dichalcogenides.  Furthermore, the fact that the model is solved in a numerically exact manner enables the treatment of high doping density.  This exact solution is facilitated by the use of GPUs, which greatly increases the efficiency of the the evaluation and manipulation of the large determinants that arise in the theory.

In making the immobile exciton approximation, the dynamics arising from this model does not conserve linear momentum.  As a consequence, the resultant 2D spectra contains an additional peak that arises from pathways that result in a singly-excited final state of the conduction electrons. We can effectively remove these pathways via a modification of the light-matter interaction.  Once so modified, the resulting 2D spectra bear a striking resemblance to that presented in Ref. \onlinecite{Tempelaar2019}, where the hole and electron masses are treated as finite.

The connection between the 2D spectra of Ref. \onlinecite{Tempelaar2019} and that of Fig. \ref{fig:fig5} provides some promising avenues for future studies.  In particular, the use of the modified light-matter interaction term in conjunction with the exactly solvable Eq. \ref{eq:MND_model} provides a route to the study of 2D spectra that removes some of the unrealistic features of the recoiless nature of our model while preserving the ability to study the high doping limit where Fermi-polaron features should be prominent.  A detailed study of this regime and the explication of the features revealed in the 2D electronic spectroscopy will be presented in future work.
\appendix
\bibliography{main}

\begin{thebibliography}{35}%
\makeatletter
\providecommand \@ifxundefined [1]{%
 \@ifx{#1\undefined}
}%
\providecommand \@ifnum [1]{%
 \ifnum #1\expandafter \@firstoftwo
 \else \expandafter \@secondoftwo
 \fi
}%
\providecommand \@ifx [1]{%
 \ifx #1\expandafter \@firstoftwo
 \else \expandafter \@secondoftwo
 \fi
}%
\providecommand \natexlab [1]{#1}%
\providecommand \enquote  [1]{``#1''}%
\providecommand \bibnamefont  [1]{#1}%
\providecommand \bibfnamefont [1]{#1}%
\providecommand \citenamefont [1]{#1}%
\providecommand \href@noop [0]{\@secondoftwo}%
\providecommand \href [0]{\begingroup \@sanitize@url \@href}%
\providecommand \@href[1]{\@@startlink{#1}\@@href}%
\providecommand \@@href[1]{\endgroup#1\@@endlink}%
\providecommand \@sanitize@url [0]{\catcode `\\12\catcode `\$12\catcode
  `\&12\catcode `\#12\catcode `\^12\catcode `\_12\catcode `\%12\relax}%
\providecommand \@@startlink[1]{}%
\providecommand \@@endlink[0]{}%
\providecommand \url  [0]{\begingroup\@sanitize@url \@url }%
\providecommand \@url [1]{\endgroup\@href {#1}{\urlprefix }}%
\providecommand \urlprefix  [0]{URL }%
\providecommand \Eprint [0]{\href }%
\providecommand \doibase [0]{https://doi.org/}%
\providecommand \selectlanguage [0]{\@gobble}%
\providecommand \bibinfo  [0]{\@secondoftwo}%
\providecommand \bibfield  [0]{\@secondoftwo}%
\providecommand \translation [1]{[#1]}%
\providecommand \BibitemOpen [0]{}%
\providecommand \bibitemStop [0]{}%
\providecommand \bibitemNoStop [0]{.\EOS\space}%
\providecommand \EOS [0]{\spacefactor3000\relax}%
\providecommand \BibitemShut  [1]{\csname bibitem#1\endcsname}%
\let\auto@bib@innerbib\@empty
\bibitem [{\citenamefont {Akinwande}\ \emph {et~al.}(2014)\citenamefont
  {Akinwande}, \citenamefont {Petrone},\ and\ \citenamefont
  {Hone}}]{Akinwande2014}%
  \BibitemOpen
  \bibfield  {author} {\bibinfo {author} {\bibfnamefont {D.}~\bibnamefont
  {Akinwande}}, \bibinfo {author} {\bibfnamefont {N.}~\bibnamefont {Petrone}},\
  and\ \bibinfo {author} {\bibfnamefont {J.}~\bibnamefont {Hone}},\ }\href
  {https://doi.org/10.1038/ncomms6678} {\bibfield  {journal} {\bibinfo
  {journal} {Nat. Commun.}\ }\textbf {\bibinfo {volume} {5}},\ \bibinfo {pages}
  {5678} (\bibinfo {year} {2014})}\BibitemShut {NoStop}%
\bibitem [{\citenamefont {Jariwala}\ \emph {et~al.}(2014)\citenamefont
  {Jariwala}, \citenamefont {Sangwan}, \citenamefont {Lauhon}, \citenamefont
  {Marks},\ and\ \citenamefont {Hersam}}]{Jariwala2014}%
  \BibitemOpen
  \bibfield  {author} {\bibinfo {author} {\bibfnamefont {D.}~\bibnamefont
  {Jariwala}}, \bibinfo {author} {\bibfnamefont {V.~K.}\ \bibnamefont
  {Sangwan}}, \bibinfo {author} {\bibfnamefont {L.~J.}\ \bibnamefont {Lauhon}},
  \bibinfo {author} {\bibfnamefont {T.~J.}\ \bibnamefont {Marks}},\ and\
  \bibinfo {author} {\bibfnamefont {M.~C.}\ \bibnamefont {Hersam}},\ }\href
  {https://doi.org/10.1021/nn500064s} {\bibfield  {journal} {\bibinfo
  {journal} {ACS Nano}\ }\textbf {\bibinfo {volume} {8}},\ \bibinfo {pages}
  {1102} (\bibinfo {year} {2014})}\BibitemShut {NoStop}%
\bibitem [{\citenamefont {Liu}\ \emph {et~al.}(2016)\citenamefont {Liu},
  \citenamefont {Weiss}, \citenamefont {Duan}, \citenamefont {Cheng},
  \citenamefont {Huang},\ and\ \citenamefont {Duan}}]{Liu2016}%
  \BibitemOpen
  \bibfield  {author} {\bibinfo {author} {\bibfnamefont {Y.}~\bibnamefont
  {Liu}}, \bibinfo {author} {\bibfnamefont {N.~O.}\ \bibnamefont {Weiss}},
  \bibinfo {author} {\bibfnamefont {X.}~\bibnamefont {Duan}}, \bibinfo {author}
  {\bibfnamefont {H.-C.}\ \bibnamefont {Cheng}}, \bibinfo {author}
  {\bibfnamefont {Y.}~\bibnamefont {Huang}},\ and\ \bibinfo {author}
  {\bibfnamefont {X.}~\bibnamefont {Duan}},\ }\href
  {https://doi.org/10.1038/natrevmats.2016.42} {\bibfield  {journal} {\bibinfo
  {journal} {Nat. Rev. Mater.}\ }\textbf {\bibinfo {volume} {1}},\ \bibinfo
  {pages} {16042} (\bibinfo {year} {2016})}\BibitemShut {NoStop}%
\bibitem [{\citenamefont {Splendiani}\ \emph {et~al.}(2010)\citenamefont
  {Splendiani}, \citenamefont {Sun}, \citenamefont {Zhang}, \citenamefont {Li},
  \citenamefont {Kim}, \citenamefont {Chim}, \citenamefont {Galli},\ and\
  \citenamefont {Wang}}]{doi:10.1021/nl903868w}%
  \BibitemOpen
  \bibfield  {author} {\bibinfo {author} {\bibfnamefont {A.}~\bibnamefont
  {Splendiani}}, \bibinfo {author} {\bibfnamefont {L.}~\bibnamefont {Sun}},
  \bibinfo {author} {\bibfnamefont {Y.}~\bibnamefont {Zhang}}, \bibinfo
  {author} {\bibfnamefont {T.}~\bibnamefont {Li}}, \bibinfo {author}
  {\bibfnamefont {J.}~\bibnamefont {Kim}}, \bibinfo {author} {\bibfnamefont
  {C.-Y.}\ \bibnamefont {Chim}}, \bibinfo {author} {\bibfnamefont
  {G.}~\bibnamefont {Galli}},\ and\ \bibinfo {author} {\bibfnamefont
  {F.}~\bibnamefont {Wang}},\ }\href {https://doi.org/10.1021/nl903868w}
  {\bibfield  {journal} {\bibinfo  {journal} {Nano Lett.}\ }\textbf {\bibinfo
  {volume} {10}},\ \bibinfo {pages} {1271} (\bibinfo {year}
  {2010})}\BibitemShut {NoStop}%
\bibitem [{\citenamefont {Mak}\ \emph {et~al.}(2010)\citenamefont {Mak},
  \citenamefont {Lee}, \citenamefont {Hone}, \citenamefont {Shan},\ and\
  \citenamefont {Heinz}}]{PhysRevLett.105.136805}%
  \BibitemOpen
  \bibfield  {author} {\bibinfo {author} {\bibfnamefont {K.~F.}\ \bibnamefont
  {Mak}}, \bibinfo {author} {\bibfnamefont {C.}~\bibnamefont {Lee}}, \bibinfo
  {author} {\bibfnamefont {J.}~\bibnamefont {Hone}}, \bibinfo {author}
  {\bibfnamefont {J.}~\bibnamefont {Shan}},\ and\ \bibinfo {author}
  {\bibfnamefont {T.~F.}\ \bibnamefont {Heinz}},\ }\href
  {https://doi.org/10.1103/PhysRevLett.105.136805} {\bibfield  {journal}
  {\bibinfo  {journal} {Phys. Rev. Lett.}\ }\textbf {\bibinfo {volume} {105}},\
  \bibinfo {pages} {136805} (\bibinfo {year} {2010})}\BibitemShut {NoStop}%
\bibitem [{\citenamefont {Mak}\ \emph {et~al.}(2012)\citenamefont {Mak},
  \citenamefont {He}, \citenamefont {Shan},\ and\ \citenamefont
  {Heinz}}]{Mak2012}%
  \BibitemOpen
  \bibfield  {author} {\bibinfo {author} {\bibfnamefont {K.~F.}\ \bibnamefont
  {Mak}}, \bibinfo {author} {\bibfnamefont {K.}~\bibnamefont {He}}, \bibinfo
  {author} {\bibfnamefont {J.}~\bibnamefont {Shan}},\ and\ \bibinfo {author}
  {\bibfnamefont {T.~F.}\ \bibnamefont {Heinz}},\ }\href
  {https://doi.org/10.1038/nnano.2012.96} {\bibfield  {journal} {\bibinfo
  {journal} {Nat. Nanotechnol.}\ }\textbf {\bibinfo {volume} {7}},\ \bibinfo
  {pages} {494} (\bibinfo {year} {2012})}\BibitemShut {NoStop}%
\bibitem [{\citenamefont {Mak}\ \emph {et~al.}(2013)\citenamefont {Mak},
  \citenamefont {He}, \citenamefont {Lee}, \citenamefont {Lee}, \citenamefont
  {Hone}, \citenamefont {Heinz},\ and\ \citenamefont {Shan}}]{Mak2013}%
  \BibitemOpen
  \bibfield  {author} {\bibinfo {author} {\bibfnamefont {K.~F.}\ \bibnamefont
  {Mak}}, \bibinfo {author} {\bibfnamefont {K.}~\bibnamefont {He}}, \bibinfo
  {author} {\bibfnamefont {C.}~\bibnamefont {Lee}}, \bibinfo {author}
  {\bibfnamefont {G.~H.}\ \bibnamefont {Lee}}, \bibinfo {author} {\bibfnamefont
  {J.}~\bibnamefont {Hone}}, \bibinfo {author} {\bibfnamefont {T.~F.}\
  \bibnamefont {Heinz}},\ and\ \bibinfo {author} {\bibfnamefont
  {J.}~\bibnamefont {Shan}},\ }\href {https://doi.org/10.1038/nmat3505}
  {\bibfield  {journal} {\bibinfo  {journal} {Nat. Mater.}\ }\textbf {\bibinfo
  {volume} {12}},\ \bibinfo {pages} {207} (\bibinfo {year} {2013})}\BibitemShut
  {NoStop}%
\bibitem [{\citenamefont {You}\ \emph {et~al.}(2015)\citenamefont {You},
  \citenamefont {Zhang}, \citenamefont {Berkelbach}, \citenamefont {Hybertsen},
  \citenamefont {Reichman},\ and\ \citenamefont {Heinz}}]{You2015}%
  \BibitemOpen
  \bibfield  {author} {\bibinfo {author} {\bibfnamefont {Y.}~\bibnamefont
  {You}}, \bibinfo {author} {\bibfnamefont {X.-X.}\ \bibnamefont {Zhang}},
  \bibinfo {author} {\bibfnamefont {T.~C.}\ \bibnamefont {Berkelbach}},
  \bibinfo {author} {\bibfnamefont {M.~S.}\ \bibnamefont {Hybertsen}}, \bibinfo
  {author} {\bibfnamefont {D.~R.}\ \bibnamefont {Reichman}},\ and\ \bibinfo
  {author} {\bibfnamefont {T.~F.}\ \bibnamefont {Heinz}},\ }\href
  {https://doi.org/10.1038/nphys3324} {\bibfield  {journal} {\bibinfo
  {journal} {Nat. Phys.}\ }\textbf {\bibinfo {volume} {11}},\ \bibinfo {pages}
  {477} (\bibinfo {year} {2015})}\BibitemShut {NoStop}%
\bibitem [{\citenamefont {Wang}\ \emph {et~al.}(2018)\citenamefont {Wang},
  \citenamefont {Chernikov}, \citenamefont {Glazov}, \citenamefont {Heinz},
  \citenamefont {Marie}, \citenamefont {Amand},\ and\ \citenamefont
  {Urbaszek}}]{RevModPhys.90.021001}%
  \BibitemOpen
  \bibfield  {author} {\bibinfo {author} {\bibfnamefont {G.}~\bibnamefont
  {Wang}}, \bibinfo {author} {\bibfnamefont {A.}~\bibnamefont {Chernikov}},
  \bibinfo {author} {\bibfnamefont {M.~M.}\ \bibnamefont {Glazov}}, \bibinfo
  {author} {\bibfnamefont {T.~F.}\ \bibnamefont {Heinz}}, \bibinfo {author}
  {\bibfnamefont {X.}~\bibnamefont {Marie}}, \bibinfo {author} {\bibfnamefont
  {T.}~\bibnamefont {Amand}},\ and\ \bibinfo {author} {\bibfnamefont
  {B.}~\bibnamefont {Urbaszek}},\ }\href
  {https://doi.org/10.1103/RevModPhys.90.021001} {\bibfield  {journal}
  {\bibinfo  {journal} {Rev. Mod. Phys.}\ }\textbf {\bibinfo {volume} {90}},\
  \bibinfo {pages} {021001} (\bibinfo {year} {2018})}\BibitemShut {NoStop}%
\bibitem [{\citenamefont {Berkelbach}\ and\ \citenamefont
  {Reichman}(2018)}]{doi:10.1146/annurev-conmatphys-033117-054009}%
  \BibitemOpen
  \bibfield  {author} {\bibinfo {author} {\bibfnamefont {T.~C.}\ \bibnamefont
  {Berkelbach}}\ and\ \bibinfo {author} {\bibfnamefont {D.~R.}\ \bibnamefont
  {Reichman}},\ }\href
  {https://doi.org/10.1146/annurev-conmatphys-033117-054009} {\bibfield
  {journal} {\bibinfo  {journal} {Annu. Rev. Condens. Matter Phys.}\ }\textbf
  {\bibinfo {volume} {9}},\ \bibinfo {pages} {379} (\bibinfo {year}
  {2018})}\BibitemShut {NoStop}%
\bibitem [{\citenamefont {Zhou}\ \emph {et~al.}(2017)\citenamefont {Zhou},
  \citenamefont {Scuri}, \citenamefont {Wild}, \citenamefont {High},
  \citenamefont {Dibos}, \citenamefont {Jauregui}, \citenamefont {Shu},
  \citenamefont {De~Greve}, \citenamefont {Pistunova}, \citenamefont {Joe},
  \citenamefont {Taniguchi}, \citenamefont {Watanabe}, \citenamefont {Kim},
  \citenamefont {Lukin},\ and\ \citenamefont {Park}}]{Zhou2017}%
  \BibitemOpen
  \bibfield  {author} {\bibinfo {author} {\bibfnamefont {Y.}~\bibnamefont
  {Zhou}}, \bibinfo {author} {\bibfnamefont {G.}~\bibnamefont {Scuri}},
  \bibinfo {author} {\bibfnamefont {D.~S.}\ \bibnamefont {Wild}}, \bibinfo
  {author} {\bibfnamefont {A.~A.}\ \bibnamefont {High}}, \bibinfo {author}
  {\bibfnamefont {A.}~\bibnamefont {Dibos}}, \bibinfo {author} {\bibfnamefont
  {L.~A.}\ \bibnamefont {Jauregui}}, \bibinfo {author} {\bibfnamefont
  {C.}~\bibnamefont {Shu}}, \bibinfo {author} {\bibfnamefont {K.}~\bibnamefont
  {De~Greve}}, \bibinfo {author} {\bibfnamefont {K.}~\bibnamefont {Pistunova}},
  \bibinfo {author} {\bibfnamefont {A.~Y.}\ \bibnamefont {Joe}}, \bibinfo
  {author} {\bibfnamefont {T.}~\bibnamefont {Taniguchi}}, \bibinfo {author}
  {\bibfnamefont {K.}~\bibnamefont {Watanabe}}, \bibinfo {author}
  {\bibfnamefont {P.}~\bibnamefont {Kim}}, \bibinfo {author} {\bibfnamefont
  {M.~D.}\ \bibnamefont {Lukin}},\ and\ \bibinfo {author} {\bibfnamefont
  {H.}~\bibnamefont {Park}},\ }\href {https://doi.org/10.1038/nnano.2017.106}
  {\bibfield  {journal} {\bibinfo  {journal} {Nat. Nanotechnol.}\ }\textbf
  {\bibinfo {volume} {12}},\ \bibinfo {pages} {856} (\bibinfo {year}
  {2017})}\BibitemShut {NoStop}%
\bibitem [{\citenamefont {Fogler}\ \emph {et~al.}(2014)\citenamefont {Fogler},
  \citenamefont {Butov},\ and\ \citenamefont {Novoselov}}]{Fogler2014}%
  \BibitemOpen
  \bibfield  {author} {\bibinfo {author} {\bibfnamefont {M.~M.}\ \bibnamefont
  {Fogler}}, \bibinfo {author} {\bibfnamefont {L.~V.}\ \bibnamefont {Butov}},\
  and\ \bibinfo {author} {\bibfnamefont {K.~S.}\ \bibnamefont {Novoselov}},\
  }\href {https://doi.org/10.1038/ncomms5555} {\bibfield  {journal} {\bibinfo
  {journal} {Nat. Commun.}\ }\textbf {\bibinfo {volume} {5}},\ \bibinfo {pages}
  {4555} (\bibinfo {year} {2014})}\BibitemShut {NoStop}%
\bibitem [{\citenamefont {Wang}\ \emph {et~al.}(2019)\citenamefont {Wang},
  \citenamefont {Rhodes}, \citenamefont {Watanabe}, \citenamefont {Taniguchi},
  \citenamefont {Hone}, \citenamefont {Shan},\ and\ \citenamefont
  {Mak}}]{Wang2019}%
  \BibitemOpen
  \bibfield  {author} {\bibinfo {author} {\bibfnamefont {Z.}~\bibnamefont
  {Wang}}, \bibinfo {author} {\bibfnamefont {D.~A.}\ \bibnamefont {Rhodes}},
  \bibinfo {author} {\bibfnamefont {K.}~\bibnamefont {Watanabe}}, \bibinfo
  {author} {\bibfnamefont {T.}~\bibnamefont {Taniguchi}}, \bibinfo {author}
  {\bibfnamefont {J.~C.}\ \bibnamefont {Hone}}, \bibinfo {author}
  {\bibfnamefont {J.}~\bibnamefont {Shan}},\ and\ \bibinfo {author}
  {\bibfnamefont {K.~F.}\ \bibnamefont {Mak}},\ }\href
  {https://doi.org/10.1038/s41586-019-1591-7} {\bibfield  {journal} {\bibinfo
  {journal} {Nature}\ }\textbf {\bibinfo {volume} {574}},\ \bibinfo {pages}
  {76} (\bibinfo {year} {2019})}\BibitemShut {NoStop}%
\bibitem [{\citenamefont {Li}\ \emph {et~al.}(2021)\citenamefont {Li},
  \citenamefont {Li}, \citenamefont {Regan}, \citenamefont {Wang},
  \citenamefont {Zhao}, \citenamefont {Kahn}, \citenamefont {Yumigeta},
  \citenamefont {Blei}, \citenamefont {Taniguchi}, \citenamefont {Watanabe},
  \citenamefont {Tongay}, \citenamefont {Zettl}, \citenamefont {Crommie},\ and\
  \citenamefont {Wang}}]{Li2021}%
  \BibitemOpen
  \bibfield  {author} {\bibinfo {author} {\bibfnamefont {H.}~\bibnamefont
  {Li}}, \bibinfo {author} {\bibfnamefont {S.}~\bibnamefont {Li}}, \bibinfo
  {author} {\bibfnamefont {E.~C.}\ \bibnamefont {Regan}}, \bibinfo {author}
  {\bibfnamefont {D.}~\bibnamefont {Wang}}, \bibinfo {author} {\bibfnamefont
  {W.}~\bibnamefont {Zhao}}, \bibinfo {author} {\bibfnamefont {S.}~\bibnamefont
  {Kahn}}, \bibinfo {author} {\bibfnamefont {K.}~\bibnamefont {Yumigeta}},
  \bibinfo {author} {\bibfnamefont {M.}~\bibnamefont {Blei}}, \bibinfo {author}
  {\bibfnamefont {T.}~\bibnamefont {Taniguchi}}, \bibinfo {author}
  {\bibfnamefont {K.}~\bibnamefont {Watanabe}}, \bibinfo {author}
  {\bibfnamefont {S.}~\bibnamefont {Tongay}}, \bibinfo {author} {\bibfnamefont
  {A.}~\bibnamefont {Zettl}}, \bibinfo {author} {\bibfnamefont {M.~F.}\
  \bibnamefont {Crommie}},\ and\ \bibinfo {author} {\bibfnamefont
  {F.}~\bibnamefont {Wang}},\ }\href
  {https://doi.org/10.1038/s41586-021-03874-9} {\bibfield  {journal} {\bibinfo
  {journal} {Nature}\ }\textbf {\bibinfo {volume} {597}},\ \bibinfo {pages}
  {650} (\bibinfo {year} {2021})}\BibitemShut {NoStop}%
\bibitem [{\citenamefont {Zhou}\ \emph {et~al.}(2021)\citenamefont {Zhou},
  \citenamefont {Sung}, \citenamefont {Brutschea}, \citenamefont {Esterlis},
  \citenamefont {Wang}, \citenamefont {Scuri}, \citenamefont {Gelly},
  \citenamefont {Heo}, \citenamefont {Taniguchi}, \citenamefont {Watanabe},
  \citenamefont {Zar{\'a}nd}, \citenamefont {Lukin}, \citenamefont {Kim},
  \citenamefont {Demler},\ and\ \citenamefont {Park}}]{Zhou2021}%
  \BibitemOpen
  \bibfield  {author} {\bibinfo {author} {\bibfnamefont {Y.}~\bibnamefont
  {Zhou}}, \bibinfo {author} {\bibfnamefont {J.}~\bibnamefont {Sung}}, \bibinfo
  {author} {\bibfnamefont {E.}~\bibnamefont {Brutschea}}, \bibinfo {author}
  {\bibfnamefont {I.}~\bibnamefont {Esterlis}}, \bibinfo {author}
  {\bibfnamefont {Y.}~\bibnamefont {Wang}}, \bibinfo {author} {\bibfnamefont
  {G.}~\bibnamefont {Scuri}}, \bibinfo {author} {\bibfnamefont {R.~J.}\
  \bibnamefont {Gelly}}, \bibinfo {author} {\bibfnamefont {H.}~\bibnamefont
  {Heo}}, \bibinfo {author} {\bibfnamefont {T.}~\bibnamefont {Taniguchi}},
  \bibinfo {author} {\bibfnamefont {K.}~\bibnamefont {Watanabe}}, \bibinfo
  {author} {\bibfnamefont {G.}~\bibnamefont {Zar{\'a}nd}}, \bibinfo {author}
  {\bibfnamefont {M.~D.}\ \bibnamefont {Lukin}}, \bibinfo {author}
  {\bibfnamefont {P.}~\bibnamefont {Kim}}, \bibinfo {author} {\bibfnamefont
  {E.}~\bibnamefont {Demler}},\ and\ \bibinfo {author} {\bibfnamefont
  {H.}~\bibnamefont {Park}},\ }\href
  {https://doi.org/10.1038/s41586-021-03560-w} {\bibfield  {journal} {\bibinfo
  {journal} {Nature}\ }\textbf {\bibinfo {volume} {595}},\ \bibinfo {pages}
  {48} (\bibinfo {year} {2021})}\BibitemShut {NoStop}%
\bibitem [{\citenamefont {Smole{\'{n}}ski}\ \emph {et~al.}(2021)\citenamefont
  {Smole{\'{n}}ski}, \citenamefont {Dolgirev}, \citenamefont {Kuhlenkamp},
  \citenamefont {Popert}, \citenamefont {Shimazaki}, \citenamefont {Back},
  \citenamefont {Lu}, \citenamefont {Kroner}, \citenamefont {Watanabe},
  \citenamefont {Taniguchi}, \citenamefont {Esterlis}, \citenamefont {Demler},\
  and\ \citenamefont {Imamo{\u{g}}lu}}]{Smolenski2021}%
  \BibitemOpen
  \bibfield  {author} {\bibinfo {author} {\bibfnamefont {T.}~\bibnamefont
  {Smole{\'{n}}ski}}, \bibinfo {author} {\bibfnamefont {P.~E.}\ \bibnamefont
  {Dolgirev}}, \bibinfo {author} {\bibfnamefont {C.}~\bibnamefont
  {Kuhlenkamp}}, \bibinfo {author} {\bibfnamefont {A.}~\bibnamefont {Popert}},
  \bibinfo {author} {\bibfnamefont {Y.}~\bibnamefont {Shimazaki}}, \bibinfo
  {author} {\bibfnamefont {P.}~\bibnamefont {Back}}, \bibinfo {author}
  {\bibfnamefont {X.}~\bibnamefont {Lu}}, \bibinfo {author} {\bibfnamefont
  {M.}~\bibnamefont {Kroner}}, \bibinfo {author} {\bibfnamefont
  {K.}~\bibnamefont {Watanabe}}, \bibinfo {author} {\bibfnamefont
  {T.}~\bibnamefont {Taniguchi}}, \bibinfo {author} {\bibfnamefont
  {I.}~\bibnamefont {Esterlis}}, \bibinfo {author} {\bibfnamefont
  {E.}~\bibnamefont {Demler}},\ and\ \bibinfo {author} {\bibfnamefont
  {A.}~\bibnamefont {Imamo{\u{g}}lu}},\ }\href
  {https://doi.org/10.1038/s41586-021-03590-4} {\bibfield  {journal} {\bibinfo
  {journal} {Nature}\ }\textbf {\bibinfo {volume} {595}},\ \bibinfo {pages}
  {53} (\bibinfo {year} {2021})}\BibitemShut {NoStop}%
\bibitem [{\citenamefont {Hao}\ \emph {et~al.}(2016)\citenamefont {Hao},
  \citenamefont {Xu}, \citenamefont {Nagler}, \citenamefont {Singh},
  \citenamefont {Tran}, \citenamefont {Dass}, \citenamefont {Sch\"uller},
  \citenamefont {Korn}, \citenamefont {Li},\ and\ \citenamefont
  {Moody}}]{doi:10.1021/acs.nanolett.6b02041}%
  \BibitemOpen
  \bibfield  {author} {\bibinfo {author} {\bibfnamefont {K.}~\bibnamefont
  {Hao}}, \bibinfo {author} {\bibfnamefont {L.}~\bibnamefont {Xu}}, \bibinfo
  {author} {\bibfnamefont {P.}~\bibnamefont {Nagler}}, \bibinfo {author}
  {\bibfnamefont {A.}~\bibnamefont {Singh}}, \bibinfo {author} {\bibfnamefont
  {K.}~\bibnamefont {Tran}}, \bibinfo {author} {\bibfnamefont {C.~K.}\
  \bibnamefont {Dass}}, \bibinfo {author} {\bibfnamefont {C.}~\bibnamefont
  {Sch\"uller}}, \bibinfo {author} {\bibfnamefont {T.}~\bibnamefont {Korn}},
  \bibinfo {author} {\bibfnamefont {X.}~\bibnamefont {Li}},\ and\ \bibinfo
  {author} {\bibfnamefont {G.}~\bibnamefont {Moody}},\ }\href
  {https://doi.org/10.1021/acs.nanolett.6b02041} {\bibfield  {journal}
  {\bibinfo  {journal} {Nano Lett.}\ }\textbf {\bibinfo {volume} {16}},\
  \bibinfo {pages} {5109} (\bibinfo {year} {2016})}\BibitemShut {NoStop}%
\bibitem [{\citenamefont {Policht}\ \emph {et~al.}(2021)\citenamefont
  {Policht}, \citenamefont {Russo}, \citenamefont {Liu}, \citenamefont
  {Trovatello}, \citenamefont {Maiuri}, \citenamefont {Bai}, \citenamefont
  {Zhu}, \citenamefont {Dal~Conte},\ and\ \citenamefont
  {Cerullo}}]{doi:10.1021/acs.nanolett.1c01098}%
  \BibitemOpen
  \bibfield  {author} {\bibinfo {author} {\bibfnamefont {V.~R.}\ \bibnamefont
  {Policht}}, \bibinfo {author} {\bibfnamefont {M.}~\bibnamefont {Russo}},
  \bibinfo {author} {\bibfnamefont {F.}~\bibnamefont {Liu}}, \bibinfo {author}
  {\bibfnamefont {C.}~\bibnamefont {Trovatello}}, \bibinfo {author}
  {\bibfnamefont {M.}~\bibnamefont {Maiuri}}, \bibinfo {author} {\bibfnamefont
  {Y.}~\bibnamefont {Bai}}, \bibinfo {author} {\bibfnamefont {X.}~\bibnamefont
  {Zhu}}, \bibinfo {author} {\bibfnamefont {S.}~\bibnamefont {Dal~Conte}},\
  and\ \bibinfo {author} {\bibfnamefont {G.}~\bibnamefont {Cerullo}},\ }\href
  {https://doi.org/10.1021/acs.nanolett.1c01098} {\bibfield  {journal}
  {\bibinfo  {journal} {Nano Lett.}\ }\textbf {\bibinfo {volume} {21}},\
  \bibinfo {pages} {4738} (\bibinfo {year} {2021})}\BibitemShut {NoStop}%
\bibitem [{\citenamefont {Purz}\ \emph {et~al.}(2021)\citenamefont {Purz},
  \citenamefont {Martin}, \citenamefont {Rivera}, \citenamefont {Holtzmann},
  \citenamefont {Xu},\ and\ \citenamefont {Cundiff}}]{PhysRevB.104.L241302}%
  \BibitemOpen
  \bibfield  {author} {\bibinfo {author} {\bibfnamefont {T.~L.}\ \bibnamefont
  {Purz}}, \bibinfo {author} {\bibfnamefont {E.~W.}\ \bibnamefont {Martin}},
  \bibinfo {author} {\bibfnamefont {P.}~\bibnamefont {Rivera}}, \bibinfo
  {author} {\bibfnamefont {W.~G.}\ \bibnamefont {Holtzmann}}, \bibinfo {author}
  {\bibfnamefont {X.}~\bibnamefont {Xu}},\ and\ \bibinfo {author}
  {\bibfnamefont {S.~T.}\ \bibnamefont {Cundiff}},\ }\href
  {https://doi.org/10.1103/PhysRevB.104.L241302} {\bibfield  {journal}
  {\bibinfo  {journal} {Phys. Rev. B}\ }\textbf {\bibinfo {volume} {104}},\
  \bibinfo {pages} {L241302} (\bibinfo {year} {2021})}\BibitemShut {NoStop}%
\bibitem [{\citenamefont {Tempelaar}\ and\ \citenamefont
  {Berkelbach}(2019)}]{Tempelaar2019}%
  \BibitemOpen
  \bibfield  {author} {\bibinfo {author} {\bibfnamefont {R.}~\bibnamefont
  {Tempelaar}}\ and\ \bibinfo {author} {\bibfnamefont {T.~C.}\ \bibnamefont
  {Berkelbach}},\ }\href {https://doi.org/10.1038/s41467-019-11497-y}
  {\bibfield  {journal} {\bibinfo  {journal} {Nat. Commun.}\ }\textbf {\bibinfo
  {volume} {10}},\ \bibinfo {pages} {3419} (\bibinfo {year}
  {2019})}\BibitemShut {NoStop}%
\bibitem [{\citenamefont {Efimkin}\ and\ \citenamefont
  {MacDonald}(2017)}]{PhysRevB.95.035417}%
  \BibitemOpen
  \bibfield  {author} {\bibinfo {author} {\bibfnamefont {D.~K.}\ \bibnamefont
  {Efimkin}}\ and\ \bibinfo {author} {\bibfnamefont {A.~H.}\ \bibnamefont
  {MacDonald}},\ }\href {https://doi.org/10.1103/PhysRevB.95.035417} {\bibfield
   {journal} {\bibinfo  {journal} {Phys. Rev. B}\ }\textbf {\bibinfo {volume}
  {95}},\ \bibinfo {pages} {035417} (\bibinfo {year} {2017})}\BibitemShut
  {NoStop}%
\bibitem [{\citenamefont {Sidler}\ \emph {et~al.}(2017)\citenamefont {Sidler},
  \citenamefont {Back}, \citenamefont {Cotlet}, \citenamefont {Srivastava},
  \citenamefont {Fink}, \citenamefont {Kroner}, \citenamefont {Demler},\ and\
  \citenamefont {Imamoglu}}]{Sidler2017}%
  \BibitemOpen
  \bibfield  {author} {\bibinfo {author} {\bibfnamefont {M.}~\bibnamefont
  {Sidler}}, \bibinfo {author} {\bibfnamefont {P.}~\bibnamefont {Back}},
  \bibinfo {author} {\bibfnamefont {O.}~\bibnamefont {Cotlet}}, \bibinfo
  {author} {\bibfnamefont {A.}~\bibnamefont {Srivastava}}, \bibinfo {author}
  {\bibfnamefont {T.}~\bibnamefont {Fink}}, \bibinfo {author} {\bibfnamefont
  {M.}~\bibnamefont {Kroner}}, \bibinfo {author} {\bibfnamefont
  {E.}~\bibnamefont {Demler}},\ and\ \bibinfo {author} {\bibfnamefont
  {A.}~\bibnamefont {Imamoglu}},\ }\href {https://doi.org/10.1038/nphys3949}
  {\bibfield  {journal} {\bibinfo  {journal} {Nat. Phys.}\ }\textbf {\bibinfo
  {volume} {13}},\ \bibinfo {pages} {255} (\bibinfo {year} {2017})}\BibitemShut
  {NoStop}%
\bibitem [{\citenamefont {Baeten}\ and\ \citenamefont
  {Wouters}(2014)}]{PhysRevB.89.245301}%
  \BibitemOpen
  \bibfield  {author} {\bibinfo {author} {\bibfnamefont {M.}~\bibnamefont
  {Baeten}}\ and\ \bibinfo {author} {\bibfnamefont {M.}~\bibnamefont
  {Wouters}},\ }\href {https://doi.org/10.1103/PhysRevB.89.245301} {\bibfield
  {journal} {\bibinfo  {journal} {Phys. Rev. B}\ }\textbf {\bibinfo {volume}
  {89}},\ \bibinfo {pages} {245301} (\bibinfo {year} {2014})}\BibitemShut
  {NoStop}%
\bibitem [{\citenamefont {Baeten}\ and\ \citenamefont
  {Wouters}(2015)}]{PhysRevB.91.115313}%
  \BibitemOpen
  \bibfield  {author} {\bibinfo {author} {\bibfnamefont {M.}~\bibnamefont
  {Baeten}}\ and\ \bibinfo {author} {\bibfnamefont {M.}~\bibnamefont
  {Wouters}},\ }\href {https://doi.org/10.1103/PhysRevB.91.115313} {\bibfield
  {journal} {\bibinfo  {journal} {Phys. Rev. B}\ }\textbf {\bibinfo {volume}
  {91}},\ \bibinfo {pages} {115313} (\bibinfo {year} {2015})}\BibitemShut
  {NoStop}%
\bibitem [{\citenamefont {Chang}\ and\ \citenamefont
  {Reichman}(2019)}]{PhysRevB.99.125421}%
  \BibitemOpen
  \bibfield  {author} {\bibinfo {author} {\bibfnamefont {Y.-W.}\ \bibnamefont
  {Chang}}\ and\ \bibinfo {author} {\bibfnamefont {D.~R.}\ \bibnamefont
  {Reichman}},\ }\href {https://doi.org/10.1103/PhysRevB.99.125421} {\bibfield
  {journal} {\bibinfo  {journal} {Phys. Rev. B}\ }\textbf {\bibinfo {volume}
  {99}},\ \bibinfo {pages} {125421} (\bibinfo {year} {2019})}\BibitemShut
  {NoStop}%
\bibitem [{\citenamefont {Nozi\`eres}\ and\ \citenamefont
  {DE~Dominicis}(1969)}]{PhysRev.178.1097}%
  \BibitemOpen
  \bibfield  {author} {\bibinfo {author} {\bibfnamefont {P.}~\bibnamefont
  {Nozi\`eres}}\ and\ \bibinfo {author} {\bibfnamefont {C.~T.}\ \bibnamefont
  {DE~Dominicis}},\ }\href {https://doi.org/10.1103/PhysRev.178.1097}
  {\bibfield  {journal} {\bibinfo  {journal} {Phys. Rev.}\ }\textbf {\bibinfo
  {volume} {178}},\ \bibinfo {pages} {1097} (\bibinfo {year}
  {1969})}\BibitemShut {NoStop}%
\bibitem [{\citenamefont {J.~Gavoret}(1969)}]{JPhysFracne.10.1051}%
  \BibitemOpen
  \bibfield  {author} {\bibinfo {author} {\bibfnamefont {B.~R. e. M.~C.}\
  \bibnamefont {J.~Gavoret}, \bibfnamefont {P.~Nozi\`eres}},\ }\href
  {https://doi.org/10.1051/jphys:019690030011-12098700} {\bibfield  {journal}
  {\bibinfo  {journal} {J. Phys.}\ }\textbf {\bibinfo {volume} {30}},\ \bibinfo
  {pages} {987} (\bibinfo {year} {1969})}\BibitemShut {NoStop}%
\bibitem [{\citenamefont {Rytova}(1967)}]{rytova1967}%
  \BibitemOpen
  \bibfield  {author} {\bibinfo {author} {\bibfnamefont {N.~S.}\ \bibnamefont
  {Rytova}},\ }\href@noop {} {\bibfield  {journal} {\bibinfo  {journal} {Vestn.
  Mosk. Univ. Fyz. Astron.}\ }\textbf {\bibinfo {volume} {3}},\ \bibinfo
  {pages} {18} (\bibinfo {year} {1967})}\BibitemShut {NoStop}%
\bibitem [{\citenamefont {Keldysh}(1979)}]{keldysh_pot}%
  \BibitemOpen
  \bibfield  {author} {\bibinfo {author} {\bibfnamefont {L.~V.}\ \bibnamefont
  {Keldysh}},\ }\href@noop {} {\bibfield  {journal} {\bibinfo  {journal} {JETP
  Lett}\ }\textbf {\bibinfo {volume} {29}},\ \bibinfo {pages} {658} (\bibinfo
  {year} {1979})}\BibitemShut {NoStop}%
\bibitem [{\citenamefont {Berkelbach}\ \emph {et~al.}(2013)\citenamefont
  {Berkelbach}, \citenamefont {Hybertsen},\ and\ \citenamefont
  {Reichman}}]{PhysRevB.88.045318}%
  \BibitemOpen
  \bibfield  {author} {\bibinfo {author} {\bibfnamefont {T.~C.}\ \bibnamefont
  {Berkelbach}}, \bibinfo {author} {\bibfnamefont {M.~S.}\ \bibnamefont
  {Hybertsen}},\ and\ \bibinfo {author} {\bibfnamefont {D.~R.}\ \bibnamefont
  {Reichman}},\ }\href {https://doi.org/10.1103/PhysRevB.88.045318} {\bibfield
  {journal} {\bibinfo  {journal} {Phys. Rev. B}\ }\textbf {\bibinfo {volume}
  {88}},\ \bibinfo {pages} {045318} (\bibinfo {year} {2013})}\BibitemShut
  {NoStop}%
\bibitem [{\citenamefont {Carbone}\ \emph {et~al.}(2020)\citenamefont
  {Carbone}, \citenamefont {Mayers},\ and\ \citenamefont
  {Reichman}}]{doi:10.1063/5.0008730}%
  \BibitemOpen
  \bibfield  {author} {\bibinfo {author} {\bibfnamefont {M.~R.}\ \bibnamefont
  {Carbone}}, \bibinfo {author} {\bibfnamefont {M.~Z.}\ \bibnamefont
  {Mayers}},\ and\ \bibinfo {author} {\bibfnamefont {D.~R.}\ \bibnamefont
  {Reichman}},\ }\href {https://doi.org/10.1063/5.0008730} {\bibfield
  {journal} {\bibinfo  {journal} {J. Chem. Phys.}\ }\textbf {\bibinfo {volume}
  {152}},\ \bibinfo {pages} {194705} (\bibinfo {year} {2020})}\BibitemShut
  {NoStop}%
\bibitem [{\citenamefont {Efimkin}\ \emph {et~al.}(2021)\citenamefont
  {Efimkin}, \citenamefont {Laird}, \citenamefont {Levinsen}, \citenamefont
  {Parish},\ and\ \citenamefont {MacDonald}}]{PhysRevB.103.075417}%
  \BibitemOpen
  \bibfield  {author} {\bibinfo {author} {\bibfnamefont {D.~K.}\ \bibnamefont
  {Efimkin}}, \bibinfo {author} {\bibfnamefont {E.~K.}\ \bibnamefont {Laird}},
  \bibinfo {author} {\bibfnamefont {J.}~\bibnamefont {Levinsen}}, \bibinfo
  {author} {\bibfnamefont {M.~M.}\ \bibnamefont {Parish}},\ and\ \bibinfo
  {author} {\bibfnamefont {A.~H.}\ \bibnamefont {MacDonald}},\ }\href
  {https://doi.org/10.1103/PhysRevB.103.075417} {\bibfield  {journal} {\bibinfo
   {journal} {Phys. Rev. B}\ }\textbf {\bibinfo {volume} {103}},\ \bibinfo
  {pages} {075417} (\bibinfo {year} {2021})}\BibitemShut {NoStop}%
\bibitem [{\citenamefont {Mukamel}(1995)}]{mukamel}%
  \BibitemOpen
  \bibfield  {author} {\bibinfo {author} {\bibfnamefont {S.}~\bibnamefont
  {Mukamel}},\ }\href@noop {} {\emph {\bibinfo {title} {Principles of Nonlinear
  Optical Spectroscopy}}}\ (\bibinfo  {publisher} {Oxford University Press},\
  \bibinfo {year} {1995})\BibitemShut {NoStop}%
\bibitem [{\citenamefont {Smallwood}\ and\ \citenamefont
  {Cundiff}(2018)}]{https://doi.org/10.1002/lpor.201800171}%
  \BibitemOpen
  \bibfield  {author} {\bibinfo {author} {\bibfnamefont {C.~L.}\ \bibnamefont
  {Smallwood}}\ and\ \bibinfo {author} {\bibfnamefont {S.~T.}\ \bibnamefont
  {Cundiff}},\ }\href {https://doi.org/https://doi.org/10.1002/lpor.201800171}
  {\bibfield  {journal} {\bibinfo  {journal} {Laser Photonics Rev.}\ }\textbf
  {\bibinfo {volume} {12}},\ \bibinfo {pages} {1800171} (\bibinfo {year}
  {2018})}\BibitemShut {NoStop}%
\bibitem [{\citenamefont {Durnev}\ and\ \citenamefont
  {Glazov}(2018)}]{Durnev_2018}%
  \BibitemOpen
  \bibfield  {author} {\bibinfo {author} {\bibfnamefont {M.~V.}\ \bibnamefont
  {Durnev}}\ and\ \bibinfo {author} {\bibfnamefont {M.~M.}\ \bibnamefont
  {Glazov}},\ }\href {https://doi.org/10.3367/ufne.2017.07.038172} {\bibfield
  {journal} {\bibinfo  {journal} {Phys.-Uspekhi}\ }\textbf {\bibinfo {volume}
  {61}},\ \bibinfo {pages} {825} (\bibinfo {year} {2018})}\BibitemShut
  {NoStop}%
\end{thebibliography}%
\begin{acknowledgements}
L.P.L. and D.R.R. were supported by the Chemical Sciences, Geosciences and Biosciences Division of the Office of Basic Energy Sciences, Office of Science, U.S. Department of Energy.  We thank Roel Tempelaar and Timothy Berkelbach for crucial discussions.
\end{acknowledgements}

\section*{Conflict of Interest}
The authors have no conflicts to disclose.

\section*{Data Availability}
The data that support the findings of this study are available from the corresponding author upon reasonable request.

\end{document}